\documentclass[twocolumn,showpacs,preprintnumbers,amsmath,amssymb,superscriptaddress]{revtex4}

\usepackage{graphicx}
\usepackage{dcolumn}
\usepackage{bm}

\newcommand{\nist}{National Institute of Standards and
Technology, Time and Frequency Division, MS 847 Boulder CO 80305}
\newcommand{\alamos}{Los Alamos National Laboratory, P-23 Physics
Division MS H803, Los Alamos NM 87545}
\newcommand{\campinas}{Gleb Wataghin Physics Institute, State
University of Campinas (UNICAMP) Brasil}

\begin{document}

\preprint{\today}

\title{Kilohertz-resolution spectroscopy of cold atoms with an optical frequency comb}

\author{T.M. Fortier}
\affiliation{\alamos}
\author{Y. Le Coq}
\affiliation{\nist}
\author{J.E. Stalnaker}
\affiliation{\nist}
\author{D.Ortega}
\affiliation{\campinas}
\author{S.A. Diddams}
\affiliation{\nist}
\author{C.W. Oates}
\affiliation{\nist}
\author{L. Hollberg}
\affiliation{\nist}

\date{\today}

\begin{abstract}
We have performed sub-Doppler spectroscopy on the narrow
intercombination line of cold calcium atoms using the amplified
output of a femtosecond laser frequency comb.  Injection locking of
a 657-nm diode laser with a femtosecond comb allows for two regimes
of amplification, one in which many lines of the comb are amplified,
and one where a single line is predominantly amplified. The output
of the laser in both regimes was used to perform kilohertz-level
spectroscopy. This experiment demonstrates the potential for
high-resolution absolute-frequency spectroscopy over the entire
spectrum of the frequency comb output using a single high-finesse
optical reference cavity.
\end{abstract}

\pacs{Valid PACS appear here}
\maketitle

The coherent optical bandwidth provided from frequency combs based
on modelocked femtosecond lasers provide the optical-to-microwave
division necessary to directly count optical frequencies, which has
lead to significant advances in optical frequency measurement
\cite{Haensch,Jones}.  With the development of ultra-broadband
titanium-doped sapphire (Ti:S) lasers \cite{Ell,Bartels,Fortier1}
and the coherent broadening possible with nonlinear optical
fibers,\cite{Ranka} the optical bandwidth of comb generators extends
from near infrared into visible frequencies. The subsequent
stabilization of such frequency combs allows for absolute optical
measurements over hundreds of terahertz making them an ideal tool
for spectroscopy. Already, frequency combs have been used for direct
spectroscopy of allowed two-photon and one-photon transitions
\cite{Haensch2,Snadden, Eckstein, Baklanov, Marian, Gerginov}.
Previous measurements with the direct output of a frequency comb
have been performed by stabilization of the comb to microwave
references and studying transitions with megahertz linewidths. A
greater potential can be realized by transferring the stability of a
narrow optical reference to the comb.

A new paradigm is attained by combining the extremely precise
optical synthesis of the comb with cold atomic samples that allow
long interaction times and permit for nonlinear spectroscopy of very
narrow optical transitions with low optical powers. Here we work
with a forbidden intercombination transition in cold neutral calcium
(Ca) and resolve features as narrow as 1.2\,kHz. In order compensate
for the relatively high residual temperatures of our atoms,we
amplifying the teeth centered at the transition wavelength in a
laser diode \cite{Cruz}. Additionally, we demonstrate the comb's
wavelength versatility by stabilizing the comb relative to a
high-finesse cavity at 534\,nm, but perform high-resolution
spectroscopy at 657\,nm. The application of direct optical
measurements with an optical frequency comb to optical clocks that
rely on both very cold atoms and optical frequency combs. Direct
spectroscopy with an optical comb could potentially yield access to
transition frequencies that would be difficult to reach otherwise.

\begin{figure}[t]
\includegraphics[width=8.5cm]{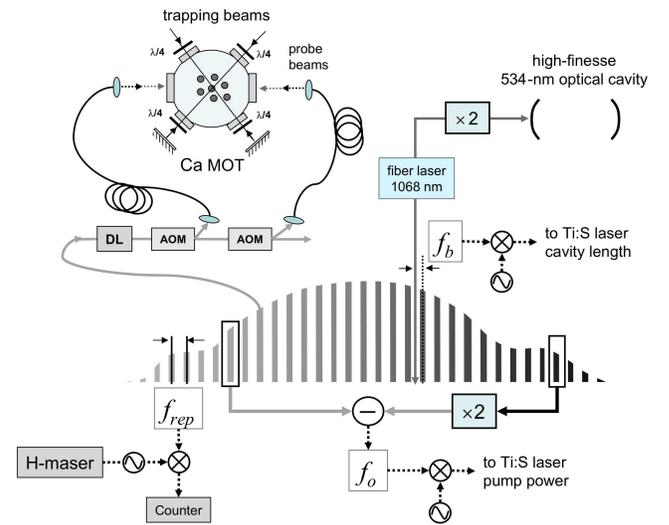}
\caption{Experimental depiction for high-precision optical
spectroscopy of cold Ca atoms using a femtosecond laser frequency
comb. Acousto-optic modulators (AOM) switch light between two
optical fibers that deliver counter-propagating pulses to the Ca
magneto-optical trap (MOT). An $f$-to-$2f$ self-referencing
technique allows phase stabilization of the laser offset frequency,
$f_0$. Phase stabilization of the laser repetition rate, $f_{\rm
rep}$, is obtained by phase locking the comb to a fiber laser at
1068 nm that is stabilized to a high-finesse optical cavity.
Measurement of the laser repetition rate versus a hydrogen-maser
permits for frequency calibration in the measurements.}
\label{FigSetup}
\end{figure}

The laser that is used in the experiments is a ring cavity
modelocked femtosecond laser based on Ti:S, with a repetition rate
around 1\,GHz \cite{Fortier2}. The laser produces an optical
spectrum with rigorously spaced and coherently related optical
frequency components that are characterized by two rf frequencies.
The first frequency is the laser repetition rate, $f_{\rm rep}$,
which sets the mode spacing and is determined by the laser cavity
length. The second is the carrier-envelope offset frequency, $f_0$,
which defines the absolute comb position and is determined by
dispersion in the laser cavity \cite{Haensch}. The laser spectrum is
composed of more than $10^5$ optical frequencies, $\nu_n$, each of
which is described absolutely by the equation, $\nu_n = n\times
f_{\rm rep} + f_0$, where $n$ is the mode number ($\sim10^5$) that
multiplies $f_{\rm rep}$ up from microwave into optical frequencies.
The laser offset frequency is stabilized using a standard
$f$-to-$2f$ interferometer, which uses a self-referencing technique
\cite{Jones} that compares frequency-doubled comb lines on the low
frequency end of the comb to fundamental light on the high frequency
end of the comb (Fig.\ \ref{FigSetup}). Stabilization of $f_0$ is
obtained by modulation of the Ti:S pump laser power (solid state
pump source at 532 nm) via an acousto-optic modulator. The octave
bandwidth necessary for stabilization of $f_0$ is obtained via
intra-cavity continuum generation in the Ti:S crystal
\cite{Fortier2}. The resulting power per mode of the Ti:S laser
spectrum is sufficient for both stabilization of the comb and and
for spectroscopy at 657 nm without additional broadening in
nonlinear fibers.

To use the modelocked laser as the local oscillator for high
precision spectroscopy requires narrow optical comb lines.  To this
end, the frequency of one mode of the comb is stabilized to a fiber
laser at 1068\,nm. Part of this fiber laser output is frequency
doubled and referenced to an optical cavity at 534 nm with a finesse
of 16,000 and a drift rate of less than 1 Hz/s \cite{Young}. Because
the optical cavity is located in a different part of the building, a
300 m long optical fiber delivers the light from the
cavity-stabilized fiber laser to the femtosecond laser frequency
comb.  The fiber length fluctuations are Doppler-canceled using
standard techniques \cite{Young2}. We obtain a heterodyne beat
signal between the comb and the fiber laser with a signal to noise
ratio of $\sim 40\,$dB in a 300\,kHz resolution bandwidth. This
radio frequency heterodyne beat is phase locked to a synthesized
reference frequency via feedback to a piezoelectric actuator that
adjusts the laser cavity length. To perform spectroscopy, we scan
the synthesized reference and hence the optical lines of the comb.
With both the frequency of a single mode and the offset frequency
stabilized, we obtain an optical linewidth of $\sim 3\,$Hz for every
comb line spanning the entire spectrum of the laser \cite{Fortier2}.
Delivery of the comb light to the Ca experiment via a 20-m long
polarization maintaining optical fiber without Doppler noise
cancelation results in a degradation of the optical linewidth to
several hundred Hertz. Noise cancelation on the fiber is imperative
for achieving sub-kilohertz resolution measurements.

\begin{figure}[htb]
\includegraphics[width=7cm]{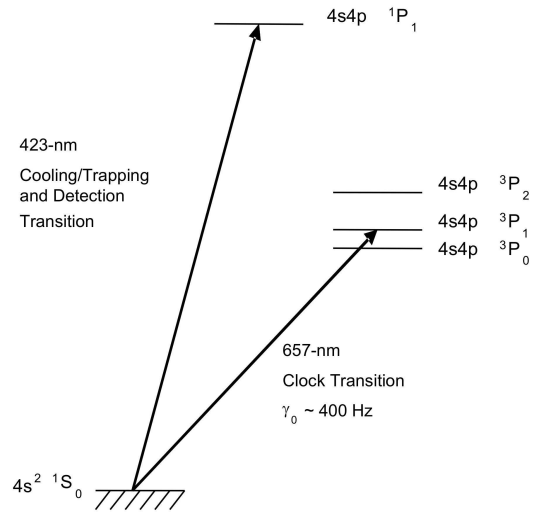} \caption{Relevant $^{40}$Ca
energy levels} \label{FigCa}
\end{figure}

We use the stabilized comb to study the narrow 657-nm
inter-combination line of the Ca atomic clock developed at NIST
\cite{Oates1}. The one-photon transition between the $4s^2$ $^1S_0$
(m = 0) and $4s4p$ $^3P_1$ (m = 0) levels, which is forbidden in the
L-S coupling approximation, has a narrow natural linewidth of
374\,Hz \cite{Degenhardt} (see Fig. \ref{FigCa}).  Using standard
trapping and cooling techniques described in detail in
\cite{Oates1}, we obtain a Ca sample with $6 \times 10^7$ atoms
cooled to 2\,mK in 2\,ms. By taking the direct output of the comb at
657 nm we have approximately 100\,nW per mode.  This power is
sufficient for measurement of the Doppler profile, but insufficient
for saturated absorption spectroscopy, which permits
higher-resolution; sub-Doppler spectroscopy of the Ca clock
transition. For higher optical powers, we use a similar technique as
that described in Ref. \cite{Cruz} by injecting comb light into an
anti-reflection coated 657-nm diode laser. The diode laser is
injected with comb light centered at 657\,nm, which is narrowed to
$\simeq 0.5\,$nm using a 2400 grooves/mm diffraction grating and a
single-mode optical fiber. For a diode laser current of 65\,mA (just
above the self-lasing threshold) this arrangement yields a uniform
ten-fold optical amplification across of the injected comb light.

\begin{figure}[t]
\includegraphics[width=8.5cm]{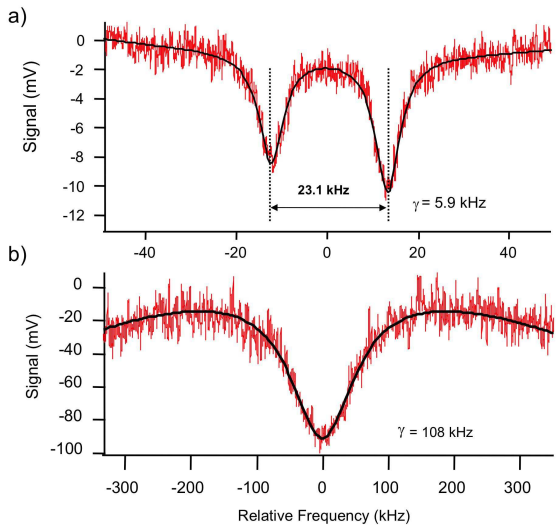}
\caption{Saturation absorption dip observed on the Doppler profile
of the Ca clock transition using two counter-propagating pulses from
a comb-injected slave laser with a current of a) 65\,mA and b)
98\,mA. The double peak observed in the case of low power broadening
(case a) is the recoil doublet.  In case b, amplification of
preferred comb lines leads to greater power broadening (linewidth =
108\,kHz), thereby making the double structure indistinguishable.
Each point in the plots is the result of 150 ms of averaging. The
y-axis is offset for convenience and 1 mV corresponds to $\sim$\,600
atoms. } \label{Fig2Pulse}
\end{figure}

The amplified comb light is delivered to the Ca atoms via optical
fiber (Fig.\ \ref{FigSetup}) whereby acousto-optic modulators act as
switches to control the time duration and the separation time of the
optical pulses delivered to the atoms (see Fig. \ref{FigSetup}).
Successive counter-propagating pulses from the comb are delivered to
the Ca atoms, which allows for a Doppler-free saturated absorption
on the optical transition. Typically, we deliver pulses with a
duration of $100\,\mu$s and an optical power of $\sim 2.7\,\mu$W
that are tuned through the resonance of the $4s^2$ $^1S_0$
$\rightarrow$ $4s4p$ $^3P_1$ transition. Excitation of the Ca sample
is measured using a shelving detection scheme
\cite{Bergquist,Oates1}, whereby fluorescence from the strongly
allowed  $^1S_0 \rightarrow ^1P_1$ transition (423\,nm), measured
before and after 657-nm excitation, reveals the ground state
depletion due to excitation to the $^3P_1$ state (see Fig.\
\ref{FigCa}). Figure \ref{Fig2Pulse}a) shows a measurement of the
sub-Doppler photon recoil splitting of 23\,kHz \cite{Uehara,Oates2}
taken with the optical frequency comb. The right peak in is the
standard saturated absorption peak (located at the center of the
Doppler profile), while the left peak is due to stimulated emission
during the second pulse and is centered at exactly one photon recoil
from the absorption peak. Note that the atomic resonance should be
exactly half way between the two peaks \cite{Oates2}. Given the
natural linewidth of the transition and the optical linewidth of the
comb light we expect a saturation linewidth on the order of 2 kHz
for a 100 $\mu$s pulse. The broader observed linewidth of 5.9\,kHz,
measured from a fit to the double-structure, results from power
broadening of the transition consistent with the applied optical
power.

Figure 3 b) shows the saturation dip when the laser diode current is
increased to 98\,mA (i.e. much higher than the threshold current).
At this diode laser current we observe preferential amplification of
particular comb teeth. By careful adjustment of the diode
temperature and current, the comb line resonant with the transition
can be made to contain up to 10 percent of the amplified power,
yielding $\sim$\,1000 times amplification of that particular mode.
This amplification factor is inferred from the power broadening
measured in the data shown in Fig.\ \ref{Fig2Pulse}b). We
independently confirmed this behavior by heterodyning the amplified
comb light against a stable CW 657\,nm optical frequency reference.
The comb light alone as measured using a fast photodetector results
in an rf spectrum with harmonics of the comb repetition rate.  The
heterodyne beat between the CW laser and the individual comb lines
is observed as sidebands on these repetition rate harmonics. At low
diode laser currents, near 65 mA, we observe sidebands with uniform
amplitude, which does not exhibit strong dependance with current and
temperature. However at 98\,mA, careful adjustment of diode laser
current allows for enhancement or suppression of the amplitude of
particular sidebands.

\begin{figure}[t]
\includegraphics[width=8.5cm]{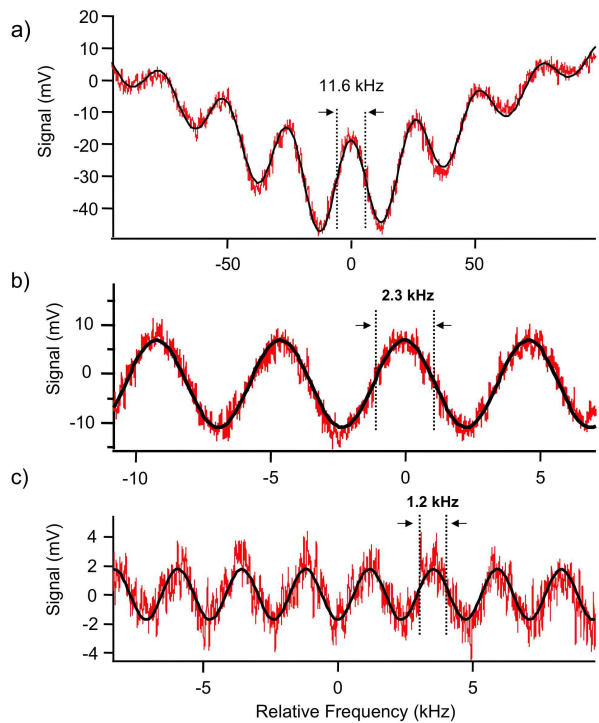}
\caption{Time-resolved optical Bord\'{e}-Ramsey fringes for pulse
lengths of $7\,\mu$s and varying pulse separations of a) $14\,\mu$s,
b) $104\,\mu$s and c) $203\,\mu$s. The decrease in the signal to
noise ratio from a) to c) is the result of spontaneous population
decay from the excited state and residual noise on the 657-nm light
(see text for details). Each point in the plots is the result of 150
ms of averaging. The y-axis is offset for convenience and 1 mV
corresponds to $\sim$\,600 atoms.} \label{Fig4Pulse}
\end{figure}

With $850\,\mu$W of amplified light at the transition wavelength we
have sufficient optical power for observation of Bord\'{e}-Ramsey
fringes, allowing for higher resolution spectroscopy of the Ca clock
transition. The technique consists of applying two $\pi/2$ pulses
from one direction (separated by a time $T/2$) and then two $\pi/2$
pulses from the opposite direction (separated by the same time
$T/2$). The excitation probability exhibits (on top of a slowly
varying envelope \cite{footnote1}) a sine wave pattern proportional
to $\sin[\pi(\nu-\nu_0)/(2\,\Delta\nu)]$, where $\Delta\nu=1/2T$ and
$\nu_0$ is the resonant optical frequency \cite{Borde}. At a diode
laser current of 65 mA the pulse duration necessary for $\pi$-pulse
would have exceeded that of the upper state lifetime. Figure
\ref{Fig4Pulse} shows the fringe contrast obtained for a fixed pulse
duration of $7\,\mu$s (required for a $\pi/2$-pulse with 850\,$\mu$W
of optical power) with varying pulse separations, $T/2$. The
relatively long pulse duration due to our limited optical power,
coupled with the broad Doppler width (~3\,MHz) on the Ca clock
transition, allows interaction with only a very narrow velocity
class of atoms. As a result only 3$\%$ of the atoms participate in
the measurement yielding a low signal to noise ratio (S/N) of $\sim
4$ after 150\,ms of averaging for our narrowest linewidth,
$\Delta\nu$ of $1.2\,$kHz. Lower residual atoms temperatures and
higher confinement would narrow the velocity distribution of the
atoms, resulting in significantly higher signal to noise ratios. The
above technique can be used for a high-resolution absolute
measurement of the line center since the AC stark shift due to other
comb components should be negligible. For a completely asymmetric
distribution of comb lines with equal amplitude, we estimate the AC
stark shift to be <\,1 mH.

In summary, we have used the amplified output of a femtosecond
optical frequency comb to demonstrated kilohertz-level spectroscopy
of the $4s4s$ $^1S_0$ (m = 0) $\rightarrow$ $4s4p$ $^3P_1$ (m = 0)
transition in a cold sample of atomic Ca. We observe two
amplification regimes when injecting a 657-nm diode laser with 0.5
nm of comb light. Near threshold, uniform amplification of the comb
light is observed, whereas at much higher diode laser currents,
preferential amplification results in a hundred-fold increase in
amplification at the transition wavelength. With our relatively high
atom temperatures, the greater amplification possible with the
second regime is necessary for higher resolution spectroscopy of the
Ca clock transition with Bord\'{e}-Ramsey fringes. Use of an atomic
sample with higher confinement and lower residual temperature,
however, could potentially allow for measurement of narrower
transitions with a single unamplified comb line.

The marriage between cold atoms and ultra-stable combs opens the
potential for using a frequency comb stabilized to a single optical
reference and performing spectroscopy at frequencies difficult to
reach otherwise. One particular example would be to explore neutral
Yb confined in an atomic lattice at NIST \cite{Barber}. The
extremely high lattice confinement and nanokelvin residual
temperatures will allow for near-unity excitation of the atoms with
a 10\,Hz resolution for less than $1\,\mu$W of optical power. Using
the unamplified output of a single comb line for probing the Yb
sample would remove the necessity to build a separate ultra-stable
probe laser at a challenging wavelength (578 nm).

\begin{acknowledgments}
This research was supported by Los Alamos National Laboratory and
the National Institute of Standards and Technology. J.E. Stalnaker
gratefully acknowledges the support of the National Research
Council. D. Ortega acknowledges the  FAPESP for support.
Additionally, we thank T. Rosenband, D. Hume and J. Bergquist for
their contributions.
\end{acknowledgments}


\begin{thebibliography}{llml}
\bibitem{footnote1} The slowly varying envelope is due to a combinaison of
the Doppler profile, finite duration of $\pi/2$ pulses, and recoil
effect \cite{Oates2}
\end{thebibliography}
\end{document}